\pdfoutput=1
\documentclass[11pt]{article}%
\usepackage{amssymb}
\usepackage[left=1.1in,right=1.1in,top=1in,bottom=1in]{geometry}
\usepackage{amsmath}
\usepackage{amsfonts}
\usepackage{graphicx}
\usepackage{natbib}
\usepackage{float}%
\providecommand{\U}[1]{\protect\rule{.1in}{.1in}}
\begin{document}

\title{Nonlocality without nonlocality\thanks{This paper is dedicated to the memory
of John A. Wheeler.}}
\author{Steven Weinstein\thanks{sw@uwaterloo.ca; sweinstein@perimeterinstitute.ca}\\Perimeter Institute for Theoretical Physics, 31 Caroline St N, Waterloo, ON
\ N2L 2Y5\\U. Waterloo Dept. of Philosophy, 200 University Ave West, Waterloo, ON\ N2L 3G1}
\date{}
\maketitle

\begin{abstract}
Bell's theorem is purported to demonstrate the impossibility of a local
\textquotedblleft hidden variable\textquotedblright\ theory underpinning
quantum mechanics. \ It relies on the well-known assumption of `locality', and
also on a little-examined assumption called `statistical
independence'\ (SI).\ Violations of this assumption have variously\ been
thought to suggest \textquotedblleft backward causation\textquotedblright, a
\textquotedblleft conspiracy\textquotedblright\ on the part of nature, or the
denial of \textquotedblleft free will\textquotedblright. \ It will be shown
here that these are spurious worries, and that denial of SI simply implies
nonlocal correlation between spacelike degrees of freedom.\ Lorentz-invariant
theories in which SI\ does not hold are easily constructed:\ two are exhibited
here.\ It is conjectured, on this basis, that quantum-mechanical phenomena may
be modeled by a local theory after all.

\end{abstract}

\section{Introduction}

The violation of the Bell-CHSH inequality by quantum mechanics is commonly
understood to undermine the possibility of \textquotedblleft
local\textquotedblright\ hidden-variable theories, i.e., theories which either
supplement quantum mechanics with additional variables (e.g., actual particle
positions in the Bohm theory \cite{Boh52a}\cite{Boh52b}), or replace the
quantum-mechanical description with something else entirely. \ This
\textquotedblleft no-go\textquotedblright\ result is known as Bell's theorem
\cite{Bell64}.\ The argument is essentially that the assumption of a certain
kind of locality -- known as `Bell locality', `factorizability'\cite{Fin80},
`strong locality'\cite{Jar84}, or simply `locality'\cite{Bell64} -- is a
sufficient condition to derive the inequality, and the predictions of quantum
mechanics violate this inequality. \ (The locution \textquotedblleft strong
locality\textquotedblright\ suggests that there are weaker forms of locality,
and indeed it was shown by Jarrett \cite{Jar84} \cite{Jar89} and by Shimony
\cite{Shim84} that `strong locality'\ is the conjunction of two weaker
locality conditions, referred to by Shimony as \textquotedblleft outcome
independence\textquotedblright\ and \textquotedblleft parameter
independence\textquotedblright.)

However, there is an additional, nontrivial assumption that goes into the Bell
argument, and this is the assumption of statistical independence (\emph{SI}).
Its role in the derivation was understood by Bell and others, but it has been
little examined because it has for the most part been thought to be beyond
question, since violations of it have seemed to most to entail either
violations of \textquotedblleft free will\textquotedblright, the possibility
of \textquotedblleft backward causation\textquotedblright, or some sort of
cosmic conspiracy on the part of nature. I will show that, to the contrary,
violations of \emph{SI} entail none of these, and I will in fact offer in
support of this contention two examples of classical, Lorentz-invariant
theories that violate \emph{SI}.

The paper proceeds as follows

\begin{itemize}
\item Section 2 rehearses the way in which the constraints of factorizability
and statistical independence come into the derivation of the Bell-CHSH inequality.

\item Section 3 shows that \emph{SI}\ entails that spacelike degrees of
freedom are independent, and that violation of \emph{SI}\ implies that
spacelike degrees of freedom are \emph{not} independent (rendering the term
\textquotedblleft degrees of freedom\textquotedblright\ something of a misnomer).

\item In section 4, it is argued that violation of \emph{SI} does not lead to
problems with free-will, backward causation, etc.

\item In section 5, two examples of \emph{SI}-violating theories are offered.

\item In section 6, it is shown that in fact violation of \emph{SI} leads
naturally to the \textquotedblleft contextuality\textquotedblright\ (of the
values of various degrees of freedom) demanded by the Kochen-Specker theorem.

\item In section 7, we discuss future prospects for a theory with nonlocal constraints.

\item Section 8 comprises some concluding remarks.
\end{itemize}

\section{Bell's theorem}

The thought experiment at the core of the Einstein-Podolsky-Rosen (EPR) paper
on the incompleteness of quantum mechanics \cite{EPR35} involves a pair of
particles in an entangled state of position and momentum, a state which is an
eigenstate of the quantum-mechanical operators representing the sum of the
momenta and the difference of the positions of the particles. Quantum
mechanics makes no definite predictions for the position and momentum of each
particle, but does make unequivocal predictions for the position or momentum
of one, given (respectively) the position or momentum of the other.
EPR\ argued that this showed that quantum mechanics must be incomplete, since
measurement of the position (or momentum) of one particle could not
simultaneously give rise to a definite position (or momentum) of the other
particle, on pain of violation of locality. They concluded that quantum
mechanics, because it did not assign a position (or momentum) to the other
particle beforehand, must be incomplete.\footnote{The argument of the
EPR\ paper is notoriously convoluted, but I follow \cite{Fin86} in regarding
this as capturing Einstein's understanding of the core argument.}

Bohm's streamlined version of the EPR\ experiment \cite{Boh51} involves the
spins of a pair of particles (either fermions or bosons) rather than their
positions and momenta.\ Prepared in what has come to be known as a
\textquotedblleft Bell\textquotedblright\ state,
\begin{equation}
\psi=\frac{1}{\sqrt{2}}(\left\vert +x\right\rangle _{A}\left\vert
-x\right\rangle _{B}-\left\vert -x\right\rangle _{A}\left\vert +x\right\rangle
_{B})\text{,}%
\end{equation}
quantum mechanics predicts that a measurement of the component of spin of
particle $A$ in any direction (e.g., $\hat{z}$) is as likely to yield $+1$ as
$-1$ (in units of $\hbar/2$), and so the expectation value $\bar{A}$ is $0$.
However, quantum mechanics also indicates that an outcome of $+1$ for a
measurement of the spin of $A$ in the $\hat{z}$ direction is guaranteed to
yield an outcome of $-1$ for $B$ for a measurement of the spin of $B$ in the
$\hat{z}$ direction. etc. This is directly analogous to the correlations
between position and momentum measurements in the original EPR experiment. \ 

In and of themselves, these phenomena offer no barrier to a hidden-variable
theory, since it is straightforward to explain such correlations by appealing
to a common cause -- the source -- and postulating that the particles emanate
from this source in (anti)correlated pairs. \ However, such an explanatory
strategy must also account for\ the way that the anticorrelation drops off as
the\ angle between the components of spin for the two particles increases
(e.g., as $A$ rotates from $\hat{x}$ toward $\hat{z}$ while $B$ remains at
$\hat{x}$). It was Bell's great insight to note that the quantum theory
implies that the anticorrelation is held onto more tightly than could be
accounted for by any \textquotedblleft local\textquotedblright\ theory. Bell
showed that the predictions of a local theory must satisfy an inequality (a
precursor to the Bell-CHSH inequality below), and that this inequality is
violated by quantum theory for appropriate choices of the components of spin
to be measured.

In order to understand the role of the locality assumption and the statistical
independence assumption, let us briefly review the derivation of the Bell-CHSH
inequality. The physical situation we are attempting to describe has the
following form:%
\[%
%TCIMACRO{\FRAME{itbpF}{3.691in}{0.7351in}{0in}{}{}{eprbox.png}%
%width 3.691in;  height 0.7351in;  depth 0in;  original-width 3.6434in;
%original-height 0.7031in;  cropleft "0";  croptop "1";  cropright "1";
%cropbottom "0";  filename 'eprbox.png';file-properties "XNPEU";}}}%
%BeginExpansion
{\includegraphics[
natheight=0.703100in,
natwidth=3.643400in,
height=0.7351in,
width=3.691in
]%
{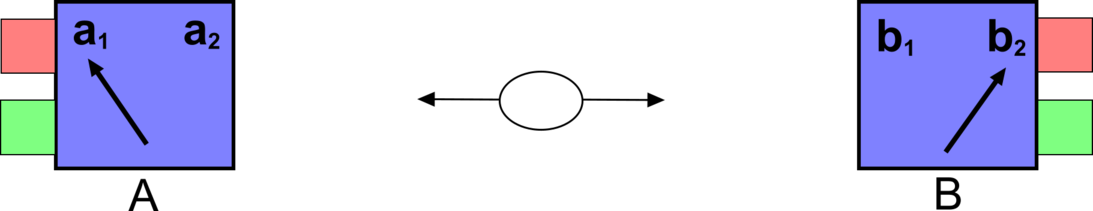}%
}%
%EndExpansion
\]
\noindent A source (represented by the ellipse) emits a pair of particles, or
in some other way causes detectors $A$ and $B$ to simultaneously (in some
frame) register one of two outcomes$.$ The detectors can be set in one of two
different ways, corresponding, in Bohm's version of the EPR\ experiment, to a
measurement of one of two different components of spin.

Let us now suppose that we have a theory that describes possible states of the
particles by a discrete or continuous parameter $\lambda$, describing either a
discrete set of states $\lambda_{1},\lambda_{2}...$ or a continuous set. \ We
will also suppose that the theory provides us with predictions for the average
value $\bar{A}(a,\lambda)$ and $\bar{B}(b,\lambda)$ of measurements of
properties $a$ and $b$ at detector $A$ and $B$ in any given state $\lambda$.
(The appeal to \emph{average} values allows for stochastic theories, in which
a given $\lambda$ might give rise to any number of different outcomes, with
various probabilities.) In general, one might suppose that $\bar{A}$ also
depended on either the detector setting $b$ or the particular outcome $B$
(i.e., $\bar{A}$ $=\bar{A}(a,\lambda,b,B)$) and similarly for $\bar{B}$.
\ That it does not, that the expectation value $\bar{A}$ in a given state
$\lambda$ does \emph{not} depend on what one chooses to measure at B, or on
the value of the distant outcome $B$ (and vice-versa) is Bell's
\emph{locality} assumption. \ Given this assumption, one can write the
expression $E(a,b,\lambda)$ for the expected product of the outcomes of
measurements of properties $a$ and $b$ in a given state $\lambda$ as
\begin{equation}
E(a,b,\lambda)=\bar{A}(a,\lambda)\bar{B}(b,\lambda)\text{.} \label{exp1}%
\end{equation}
This condition is also known as `factorizability', deriving as it does from
the fact that the joint probability of a pair of outcomes can be factorized
into the product of the marginal probabilities of each outcome. We can thus
represent the analysis of the experimental arrangement in this way:%
\[%
%TCIMACRO{\FRAME{itbpF}{4.606in}{0.8691in}{0in}{}{}{eprbox12.png}%
%width 4.606in;  height 0.8691in;  depth 0in;  original-width 4.5532in;
%original-height 0.8371in;  cropleft "0";  croptop "1";  cropright "1";
%cropbottom "0";  filename 'eprbox12.png';file-properties "XNPEU";}}}%
%BeginExpansion
{\includegraphics[
natheight=0.837100in,
natwidth=4.553200in,
height=0.8691in,
width=4.606in
]%
{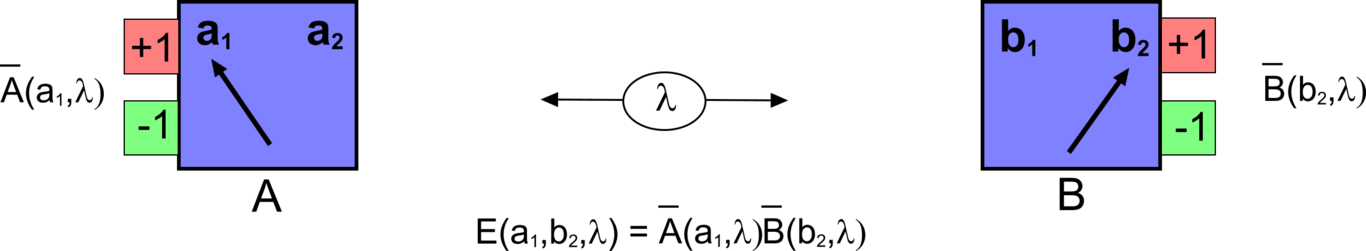}%
}%
%EndExpansion
\]

Now, a theory that accounts for our observations will presumably do so in part
by giving a probability distribution $P(\lambda)$ over the various possible
states associated with a given \textquotedblleft preparation\textquotedblright%
\ (a given set of circumstances in the region of the ellipse in the diagram
above), and the expected outcome $E(a,b)$ will then be given by the weighted
sum (we restrict to discrete $\lambda~$for simplicity)
\begin{equation}
E(a,b)=%
%TCIMACRO{\dsum \limits_{\lambda}}%
%BeginExpansion
{\displaystyle\sum\limits_{\lambda}}
%EndExpansion
E(a,b,\lambda)P(\lambda|a,b)=%
%TCIMACRO{\dsum \limits_{\lambda}}%
%BeginExpansion
{\displaystyle\sum\limits_{\lambda}}
%EndExpansion
\bar{A}(a,\lambda)\bar{B}(b,\lambda)P(\lambda|a,b) \label{exp2}%
\end{equation}
where $P(\lambda|a,b)$ is the probability of $\lambda~$given detector settings
$a$ and $b$. Thus the expected value for the product of a measurement of spin
components $a_{1}$ and $b_{2}$ is
\begin{equation}
E(a_{1},b_{2})=%
%TCIMACRO{\dsum \limits_{\lambda}}%
%BeginExpansion
{\displaystyle\sum\limits_{\lambda}}
%EndExpansion
\bar{A}(a_{1},\lambda)\bar{B}(b_{2},\lambda)P(\lambda|a_{1},b_{2})\text{ ,}
\label{exp3}%
\end{equation}
the sum of the products of the expected values of the outcome at A, the
outcome at B in each state $\lambda$ ($\lambda_{1},\lambda_{2}$, etc.), and
the probability of that state. \emph{If} the probability of $\lambda$ is
\emph{independent} of the detector settings $a$ and $b$, then one can replace
$P(\lambda|a,b)$ with $P(\lambda)$. This is the condition known as
\emph{Statistical Independence} (\emph{SI}), and as we shall now see, it is
crucial to the derivation of Bell's result

The Bell-CHSH inequality \cite{CHSH69}, is:%

\begin{equation}
\left\vert E(a_{1},b_{1})-E(a_{1},b_{2})\right\vert +\left\vert E(a_{2}%
,b_{2})+E(a_{2},b_{1})\right\vert \leq2\text{ .} \label{CHSH}%
\end{equation}
The beginning of the derivation goes as follows. \ First, write down the
difference between expectation values for pairs of settings $\left\langle
a_{1},b_{1}\right\rangle $ and $\left\langle a_{1},b_{2}\right\rangle :$%
\begin{equation}%
\begin{array}
[c]{ll}%
E(a_{1},b_{1})-E(a_{1},b_{2}) & =%
%TCIMACRO{\dsum \limits_{\lambda}}%
%BeginExpansion
{\displaystyle\sum\limits_{\lambda}}
%EndExpansion
\bar{A}(a_{1},\lambda)\bar{B}(b_{1},\lambda)P(\lambda|a_{1},b_{1})-%
%TCIMACRO{\dsum \limits_{\lambda}}%
%BeginExpansion
{\displaystyle\sum\limits_{\lambda}}
%EndExpansion
\bar{A}(a_{1},\lambda)\bar{B}(b_{2},\lambda)P(\lambda|a_{1},b_{2})
\end{array}
\label{CHSH1}%
\end{equation}
Assuming that \emph{SI} holds, we can rewrite this as%
\begin{equation}%
\begin{array}
[c]{ll}%
E(a_{1},b_{1})-E(a_{1},b_{2}) & =%
%TCIMACRO{\dsum \limits_{\lambda}}%
%BeginExpansion
{\displaystyle\sum\limits_{\lambda}}
%EndExpansion
\bar{A}(a_{1},\lambda)\bar{B}(b_{1},\lambda)P(\lambda)-%
%TCIMACRO{\dsum \limits_{\lambda}}%
%BeginExpansion
{\displaystyle\sum\limits_{\lambda}}
%EndExpansion
\bar{A}(a_{1},\lambda)\bar{B}(b_{2},\lambda)P(\lambda)
\end{array}
\label{CHSH1a}%
\end{equation}
The key step, which allows the introduction of $E(a_{2},b_{2})$ and
$E(a_{2},b_{1})$, involves expanding this as
\begin{equation}
E(a_{1},b_{1})-E(a_{1},b_{2})=%
\begin{array}
[c]{c}%
%TCIMACRO{\dsum \limits_{\lambda}}%
%BeginExpansion
{\displaystyle\sum\limits_{\lambda}}
%EndExpansion
\bar{A}(a_{1},\lambda)\bar{B}(b_{1},\lambda)P(\lambda)(1\pm\bar{A}%
(a_{2},\lambda)\bar{B}(b_{2},\lambda))\\
-%
%TCIMACRO{\dsum \limits_{\lambda}}%
%BeginExpansion
{\displaystyle\sum\limits_{\lambda}}
%EndExpansion
\bar{A}(a_{1},\lambda)\bar{B}(b_{2},\lambda)P(\lambda)(1\pm\bar{A}%
(a_{2},\lambda)\bar{B}(b_{1},\lambda))\text{ .}%
\end{array}
\label{CHSH1b}%
\end{equation}
This then leads to (\ref{CHSH}) via rearrangement of terms and manipulation
using the relations $\left\vert x\right\vert \left\vert y\right\vert
=\left\vert xy\right\vert $ and $\left\vert x+y\right\vert \leq\left\vert
x\right\vert +\left\vert y\right\vert $. For our purposes, though, the crucial
step is (\ref{CHSH1a}), in which essential use is made of \emph{SI}. If we
were not to assume \emph{SI}, then (\ref{CHSH1a}) would revert to
(\ref{CHSH1}) and we would have to rewrite (\ref{CHSH1b}) as
\begin{equation}
E(a_{1},b_{1})-E(a_{1},b_{2})%
%TCIMACRO{\TeXButton{TeX field}{\stackrel{_?}{=}}}%
%BeginExpansion
\stackrel{_?}{=}%
%EndExpansion%
\begin{array}
[c]{c}%
%TCIMACRO{\dsum \limits_{\lambda}}%
%BeginExpansion
{\displaystyle\sum\limits_{\lambda}}
%EndExpansion
\bar{A}(a_{1},\lambda)\bar{B}(b_{1},\lambda)P(\lambda|a_{1},b_{1})(1\pm\bar
{A}(a_{2},\lambda)\bar{B}(b_{2},\lambda))\\
-%
%TCIMACRO{\dsum \limits_{\lambda}}%
%BeginExpansion
{\displaystyle\sum\limits_{\lambda}}
%EndExpansion
\bar{A}(a_{1},\lambda)\bar{B}(b_{2},\lambda)P(\lambda|a_{1},b_{2})(1\pm\bar
{A}(a_{2},\lambda)\bar{B}(b_{1},\lambda))
\end{array}
\label{CHSH2b}%
\end{equation}
which is simply \emph{invalid}, since the new terms need not sum to zero
anymore. Nor, for that matter, would they correspond to the desired
$E(a_{2},b_{2})$ and $E(a_{2},b_{1})$. Thus without appeal to \emph{SI}, there
is no way to introduce the other two expectation values and derive the inequality.

\section{Statistical independence revisited}

The assumption of \emph{SI}\ has been called into question only infrequently,
but when it has, the critique has often been motivated by an appeal to the
plausibility of Lorentz-invariant \textquotedblleft backward
causation\textquotedblright, whereby the change of detector settings gives
rise to effects which propagate along or within the backward lightcone and
thereby give rise to nontrivial initial correlations in the particle
properties encoded in $\lambda$ (e.g., \cite{Cos78},\cite{Suth83},
\cite{Pri96}). \ In this section, I will argue that this is an inappropriate
way to motivate the rejection of \emph{SI}, and that its rejection instead
involves a relativistically nonproblematic commitment to nonlocal constraints
on initial data.

\begin{figure}
[ptb]
\begin{center}
\includegraphics[
natheight=2.889300in,
natwidth=3.693600in,
height=2.9334in,
width=3.7412in
]%
{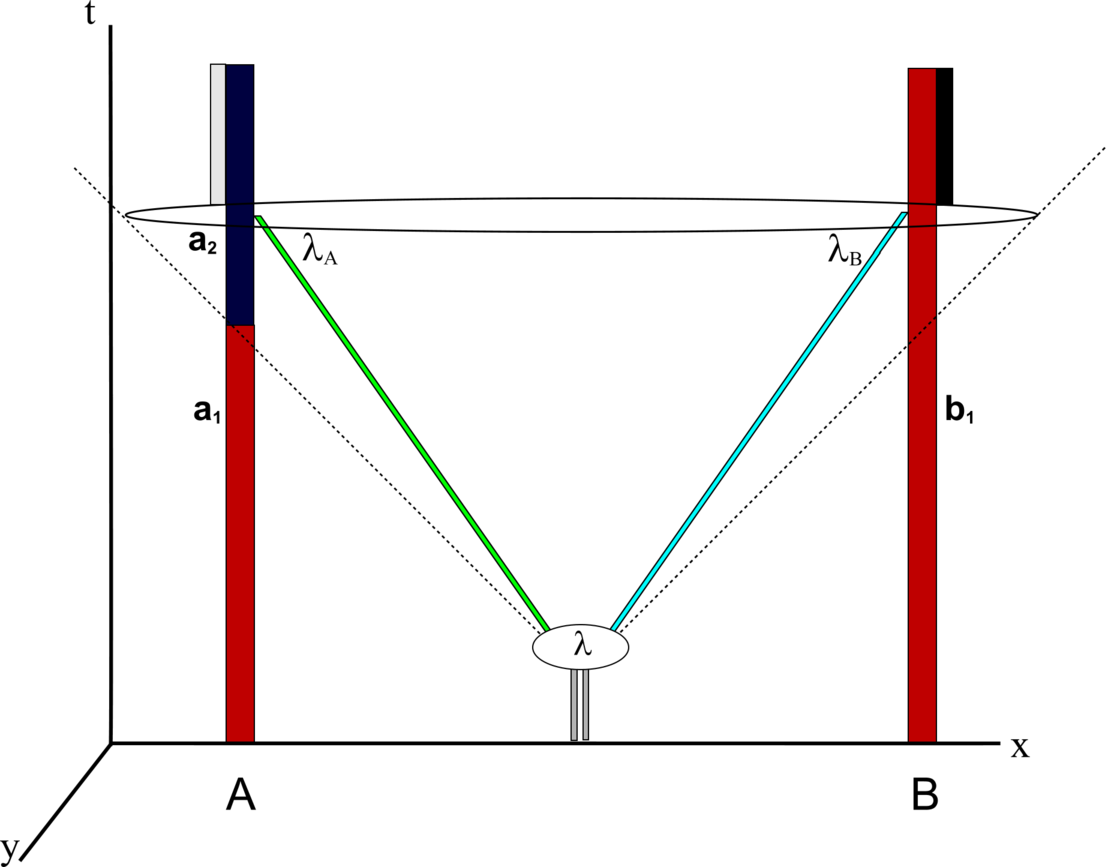}%
\caption{EPR:\ Spacetime diagram}%
\label{EPRfw}%
\end{center}
\end{figure}

Depicted in Figure \ref{EPRfw} is a run in which the setting of A is changed
from $a_{1}$ to $a_{2}$ while the particles (or whatever it is that emanates
from the source) are in flight.%

%TCIMACRO{\FRAME{ftbpFU}{3.7412in}{2.9334in}{0pt}{\Qcb{EPR:\ Spacetime
%diagram}}{\Qlb{EPRfw}}{epr3.png}{\special{ language "Scientific Word";
%type "GRAPHIC";  maintain-aspect-ratio TRUE;  display "USEDEF";
%valid_file "F";  width 3.7412in;  height 2.9334in;  depth 0pt;
%original-width 3.6936in;  original-height 2.8893in;  cropleft "0";
%filename 'epr3.png';file-properties "XNPEU";}}}%
%BeginExpansion
%EndExpansion
The different colors subsequent to the arrival of the particles at the two
detectors correspond to distinct experimental outcomes.

\begin{figure}
[ptb]
\begin{center}
\includegraphics[
natheight=2.693900in,
natwidth=4.472800in,
height=2.7363in,
width=4.5256in
]%
{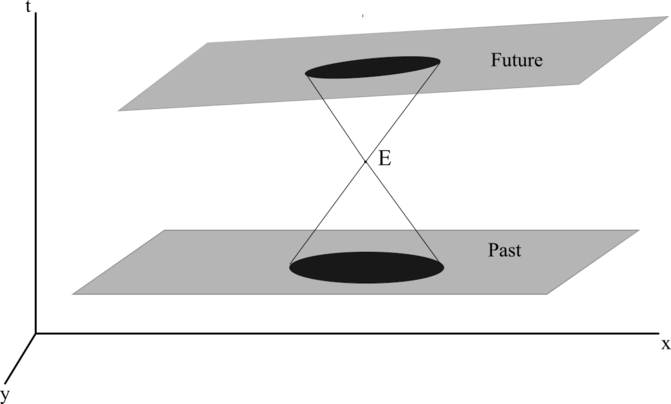}%
\caption{Past and future domains of dependence}%
\label{Cauchy}%
\end{center}
\end{figure}

Now a clear-thinking student of relativity should suspect that something is
amiss with this argument, since all deterministic theories in use today
already sanction a \emph{form} of backward causation, in that they allow both
prediction and retrodiction.\ All special and general-relativistic theories
have a well-defined causal structure which makes no distinction, other than a
conventional one, between future and past. Specifying the physical properties
(the Cauchy data) at each point on either of the shaded surfaces suffices to
determine the physical situation at $E$ (Figure \ref{Cauchy}).%
%TCIMACRO{\FRAME{ftbpFU}{4.5256in}{2.7363in}{0pt}{\Qcb{Past and future domains
%of dependence}}{\Qlb{Cauchy}}{spacelike.png}%
%width 4.5256in;  height 2.7363in;  depth 0pt;  original-width 4.4728in;
%original-height 2.6939in;  cropleft "0";  croptop "1";  cropright "1";
%cropbottom "0";  filename 'spacelike.png';file-properties "XNPEU";}}}%
%BeginExpansion
%EndExpansion
The future data determine the event $E$ just as much as the past data. And
given an appropriate description of the future data---a description which
adopts a \textquotedblleft backward-directed\textquotedblright\ temporal
orientation---one can regard these data as the \emph{cause} of the event
$E$.\footnote{For example, we seek the cause of an explosion in the past, but
if the very same event were described from an inverted temporal perspective,
described as an implosion, we would look in the opposite temporal direction.}

\begin{figure}
[ptb]
\begin{center}
\includegraphics[
natheight=2.889300in,
natwidth=3.693600in,
height=2.9334in,
width=3.7412in
]%
{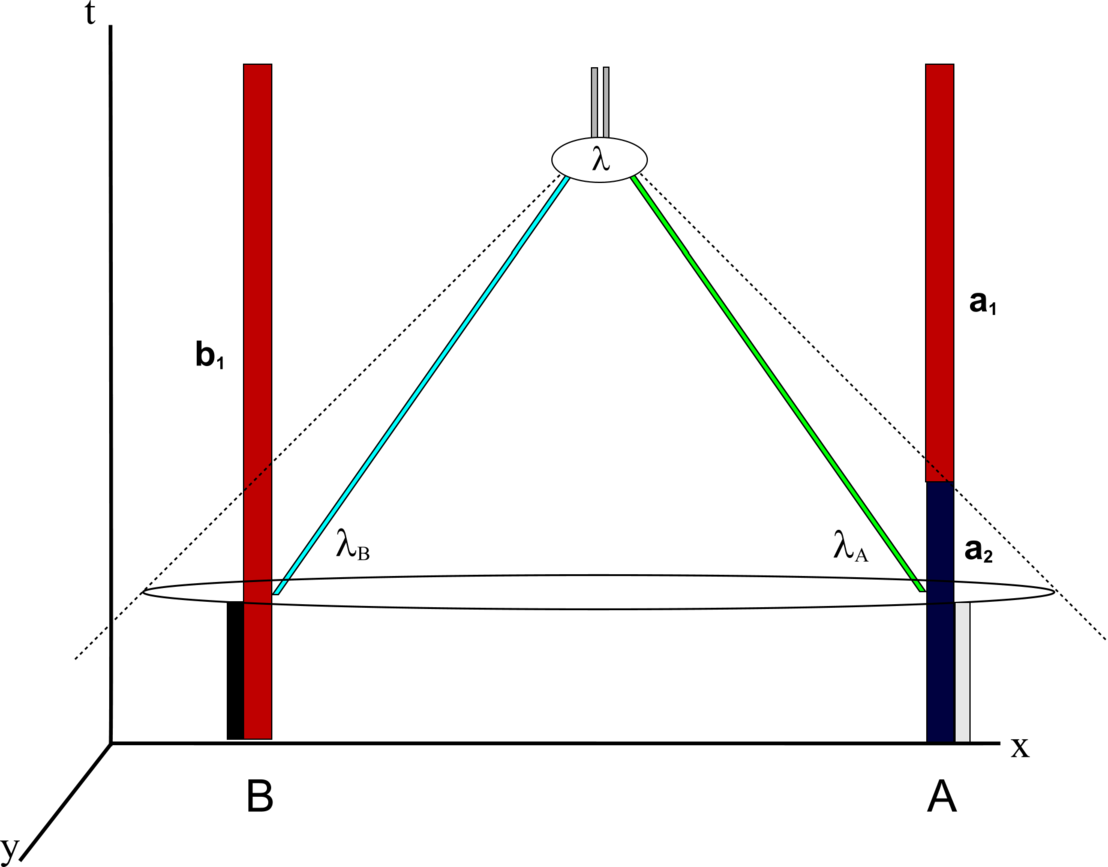}%
\caption{EPR:\ Spacetime diagram with inverted time axis}%
\label{EPRflip}%
\end{center}
\end{figure}

Let us put aside any qualms we might have regarding the notion of backward
causation for the moment and examine the particular situation of the EPR-Bohm
experiment more closely, in the hope that this will shed some light. Suppose
we simply temporally invert the situation above, as in Figure \ref{EPRflip}.%
%TCIMACRO{\FRAME{ftbpFU}{3.7412in}{2.9334in}{0pt}{\Qcb{EPR:\ Spacetime diagram
%with inverted time axis}}{\Qlb{EPRflip}}{epr3r.png}%
%width 3.7412in;  height 2.9334in;  depth 0pt;  original-width 3.6936in;
%original-height 2.8893in;  cropleft "0";  croptop "1";  cropright "1";
%cropbottom "0";  filename 'epr3r.png';file-properties "XNPEU";}}}%
%BeginExpansion
%EndExpansion
This looks like a pair of sources, $A$ and $B$, emitting particles in the
direction of a common destination. \ Is it not reasonable to expect that the
\textquotedblleft final\textquotedblright\ state $\lambda=(\lambda_{A}%
$,$\lambda_{B})$ is correlated with the settings of the sources?

In fact, it is not. Suppose we know some, but not all, of the data in the past
lightcone of an event $E$. Suppose, for instance, that we know that the white
region is empty of physically significant data, and suppose we know the data
in the red and blue ellipses, but not in the turquoise ellipses indicated by question-marks:%

\[%
%TCIMACRO{\FRAME{itbpF}{2.0124in}{1.7028in}{0in}{}{\Qlb{Par1}}{spacepar1.png}%
%width 2.0124in;  height 1.7028in;  depth 0in;  original-width 1.9735in;
%original-height 1.6665in;  cropleft "0";  croptop "1";  cropright "1";
%cropbottom "0";  filename 'spacepar1.png';file-properties "XNPEU";}}}%
%BeginExpansion
{\includegraphics[
natheight=1.666500in,
natwidth=1.973500in,
height=1.7028in,
width=2.0124in
]%
{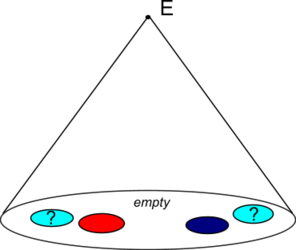}%
}%
%EndExpansion
\]
In such a case, we know nothing more about what to expect at $E$ than if we
had no information at all. Given the ability to choose data in the turquoise
ellipsoids, we can make $E$ whatever we want.

But now suppose we fill in the unspecified ellipses:%

\[%
%TCIMACRO{\FRAME{itbpF}{2.0124in}{1.7028in}{0in}{}{\Qlb{Par2}}{spacepar2.png}%
%width 2.0124in;  height 1.7028in;  depth 0in;  original-width 1.9735in;
%original-height 1.6665in;  cropleft "0";  croptop "1";  cropright "1";
%cropbottom "0";  filename 'spacepar2.png';file-properties "XNPEU";}}}%
%BeginExpansion
{\includegraphics[
natheight=1.666500in,
natwidth=1.973500in,
height=1.7028in,
width=2.0124in
]%
{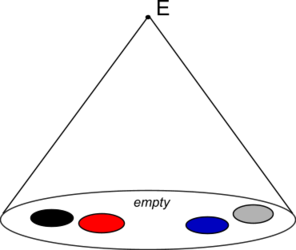}%
}%
%EndExpansion
\]
\noindent Then we know data on a Cauchy slice of the past lightcone of $E$,
and $E$ is fully determined. \ It is \emph{this} situation that is the
appropriate parallel of the time-reversed EPR experiment; the newly-specified
data correspond to the outcomes of the two trials.

Any plausibility that the particles might be \emph{causally }correlated with
the detector settings derives from a situation in which the detectors
themselves are the sources of the particles, rather than mere conduits. In the
EPR case, there are meaningful, (anti)correlated detection events, and in the
time-reversed picture these detection events serve as additional data,
additional sources. \ (One can move the slice so that it is prior to any
outcome, but one will still have to contend with the fact that only a complete
specification of the physics inside the detector, \emph{including the state of
the particles}, is sufficient to determine $E$.)

The upshot, then, is that postulating a correlation between detector settings
and the initial state $\lambda$\ (corresponding to $E$ in the figure above) --
i.e., dropping \emph{SI} -- amounts to postulating a correlation between the
detector settings and the particle properties at \emph{any} given time. I.e.,
the correlations are not causal -- \emph{they are not brought about
dynamically} -- but are properties of the state at \emph{any }instant. The
role of dynamics in a proper theory describing quantum phenomena is to
\emph{enforce}, not \emph{generate}, such correlations. The challenge, as yet
unmet, is to articulate the constraint which encodes these correlations. In
section 6, we will examine two theories with appropriately nonlocal
constraints\ in order to develop intuition. But first, let us examine two
other problems which are purported to arise in the rejection of \emph{SI}

\section{Superdeterminism and free will}

The idea that the rejection of \emph{SI} involves not some dynamical process
like\ \textquotedblleft backward causation\textquotedblright\ but rather some
preexisting and persisting\ correlations between subsystems has been broached
before, under the terms `conspiracy theory', `hyperdeterminism', and
`superdeterminism'. Bell \cite{Bel90b}, Shimony \cite{Shim85}, Lewis
\cite{LewP06} and others have suggested that proposing a correlation between
detector settings and particle properties involves some sort of conspiracy on
the part of nature. This is frequently accompanied by the charge that the
existence of such correlations is a threat to \textquotedblleft free
will\textquotedblright. Let us address these worries.

\subsection{Conspiracy}

The idea that postulating a correlation between detector settings and particle
properties involves a \textquotedblleft conspiracy\textquotedblright\ on the
part of nature appears to derive from the idea that it amounts to postulating
that the initial conditions of nature have been set up in anticipation of our
measurements. It might be supposed, analogously, that every time I telephone
my friend Jenny at 867-5309, something -- perhaps a cosmic ray -- causes my
message to be misdirected to the non-working number 867-5308, so that I appear
to live in a Kafkaesque world in which my efforts to contact Jenny are forever
stymied. This would appear to be a world in which nature, in the form of
particularly vicious initial conditions, conspires against me to the point
where I am driven to postulate that it is a law of nature that I cannot
successfully contact Jenny (except, perhaps, on her mobile phone). But
according to the way the story is told, it is really just an accident that I
cannot successfully make contact. Similarly, the conspiracy theorist views the
appeal to a failure of \emph{SI} in order to explain the strange correlations
predicted by quantum theory as an appeal to a vast conspiracy on the part of
nature to set initial conditions in such a way as to ensure that experiments
come out in accord with the quantum-mechanical predictions, so that every time
I do an EPR\ experiment it just happens to be the case that the detectors are
set in a way appropriate to generate the observed correlations.

What the conspiracy theorist is in effect doing is supposing a non-lawlike
suspension of \emph{SI}. \ That is, she is supposing that the laws of nature
are ordinary, local, relativistic laws, \emph{without} any nonlocal
constraints, but that the initial conditions are such that it happens to turn
out that the states of measuring apparatuses are nontrivially, and
persistently, correlated with the states of the particles they eventually
interact with.\ The idea seems to be that, were the initial conditions to have
been somewhat different, the entire quantum-mechanical edifice would fall
apart. Certainly, this is a theoretical possibility, but not a very happy one,
for two reasons. Were one to maintain that the laws of nature are the ordinary
classical ones, with no general, nonlocal constraints, and that quantum
mechanics is the result of a highly special set of initial conditions, one
would be foregoing the possibility of explaining the myriad phenomena
accounted for by quantum mechanics which have \emph{nothing to do }with
measurement, such as the stability of matter or the black-body emission
spectrum. Although a theory that purports to account for the full spectrum of
quantum phenomena in a way that does not violate Bell's locality assumptions
must specify nontrivial correlations between spacelike degrees of freedom, it
cannot do \emph{just} that. \ Rather, the constraints in an \emph{SI}%
-violating theory must account for the full range of phenomena accounted for
by quantum mechanics and quantum field theory. Thus a truly useful and
predictive theory underpinning quantum phenomena is highly unlikely to have
the \emph{ad hoc} character which concerns the conspiracy theorist.

\subsection{Free will}

Another worry about giving up \emph{SI} and postulating generic nonlocal,
spacelike correlations has to do with a purported threat to our
\textquotedblleft free will\textquotedblright. This particular concern has
been the subject of renewed debate in the last couple of years, prompted in
part by an argument of Conway \&\ Kochen \cite{CK06}. \ The core of the worry
is that if detector settings are correlated with particle properties, this
must mean that we cannot \textquotedblleft freely choose\textquotedblright%
\ the detector settings. This worry, however, appears to be based on a
conception of free will which is incompatible with ordinary determinism, as
pointed out by 't Hooft \cite{tHo07}. Why is it any more of a threat to free
will to have our \textquotedblleft actions\textquotedblright\ correlated with
other degrees of freedom than it is to have our actions be \emph{determined}
by the events in our past? \ (Conway and Kochen bite the bullet and argue that
even ordinary determinism is incompatible with free will.)

One might conceivably make the case that ordinary deterministic theories are
fine in a way that superdeterministic theories, with their nonlocal
constraints, are not by arguing that, in allowing that our actions are
determined by the past, we are simply granting that our actions arise from our
own thoughts and inclinations. This (limited) sort of determination is
actually \emph{essential }for free choice. This perspective is a version of
what is called `compatibilism' in philosophy, the view that freedom of the
will is compatible with determinism \cite{Hume00}.

A problem for free will would then arise if it were the case that the nonlocal
correlations associated with an underlying superdeterministic theory somehow
\emph{prevented} an agent from acting on its thoughts and inclinations. This
is to say that the physical object identified as the \textquotedblleft
agent\textquotedblright\ would exhibit behavior not explicable in terms of the
influences on or in its past lightcone. \ But this is not what is being
contemplated here, for this would involve non-Lorentz-invariant dynamics. What
is at issue are Lorentz-invariant theories with \emph{nonlocal constraints}.

\section{Theories with nonlocal constraints: two examples}

Theories which have constraints on initial data can be divided into two kinds,
local and nonlocal. \ The gauge field theories of the standard model are
local, in that the constraint may be expressed as a local condition, for
example $\nabla\cdot E=0$, the Gauss law (in \emph{vacuo}). \ The locality of
the constraint means that specifying the field at every point outside of an
open set surrounding a point $x$ does not constrain the value of the field at
$x$. Rather, the field in the neighborhood of a point is constrained only by
the field \emph{at} the point.

Theories with nonlocal constraints are less familiar. \ These are theories in
which specifying the value of a field outside the neighborhood of a point $x$
constrains the field at $x$. We will now consider two examples of such theories.

\subsection{Timelike Cauchy surfaces}

Consider the theory of the massless scalar field, given by the wave equation
$\square\phi=0$. \ In two space dimensions, this reads%
\begin{equation}
\left(  \frac{\partial^{2}}{\partial x_{1}^{2}}+\frac{\partial^{2}}{\partial
x_{2}^{2}}-\frac{1}{c^{2}}\frac{\partial^{2}}{\partial t^{2}}\right)  \phi=0
\end{equation}
where $\phi(x,t)$ is a twice-differentiable, real-valued field on
spacetime.\ It is well-known that the Cauchy problem is well-posed, meaning
that specifying the field $\phi(x)$ and its normal derivative $\partial
\phi(x)/\partial t$ on a spacelike hyperplane $t=\emph{0}$ uniquely fixes the
field at all other times. More important, it is also the case that a solution
exists for any such data, meaning that the field and its time rate-of-change
at each point are independently specifiable. Thus the ordinary initial value
formulation of the wave equation has no constraints, either local or nonlocal.

On the other hand one can also specify initial data on a \emph{mixed}
(spacelike and timelike) hyperplane \cite{Joh91}.\footnote{These mixed
hypersurfaces are sometimes called \textquotedblleft
timelike\textquotedblright\ \cite{Joh91} or \textquotedblleft
non-spacelike\textquotedblright\ \cite{Cou62}.} Given data $\phi(x_{1},t)$ and
$\partial\phi(x_{1},t)/\partial x_{2}$ on the hyperplane $x_{2}=0$, the data
uniquely determine the solution, if a solution exists at all for that data.
The fact that solutions do not exist for arbitrary data (except in the case of
one space dimension) means that, as formulated, the initial value problem is
not \textquotedblleft well-posed.\textquotedblright\ However, it has recently
been shown \cite{CW08} that, just as in a gauge theory, one may write down a
constraint on the initial data such that \emph{any} initial data satisfying
the constraint lead to a unique, stable solution of the equation. The
resulting problem \emph{is} well-posed.

The difference between this constraint and the gauge-theoretic constraint is
that the former is nonlocal while the latter is local. \ Specifically, the
Cauchy data is given by functions $f$ and $g$ such that
\begin{align}
f(x_{1},t)  &  :=\phi(x_{1},0,t)=%
%TCIMACRO{\dint \limits_{k_{1}^{2}\geq\omega^{2}}}%
%BeginExpansion
{\displaystyle\int\limits_{k_{1}^{2}\geq\omega^{2}}}
%EndExpansion
\tilde{f}(k_{1},\omega)e^{i(k_{1}x_{1}-\omega t)}dk_{1}d\omega\\
g(x_{1},t)  &  :=\frac{\partial\phi(x_{1},0,t)}{\partial x_{2}}=%
%TCIMACRO{\dint \limits_{k_{1}^{2}\geq\omega^{2}}}%
%BeginExpansion
{\displaystyle\int\limits_{k_{1}^{2}\geq\omega^{2}}}
%EndExpansion
\tilde{g}(k_{1},\omega)e^{i(k_{1}x_{1}-\omega t)}dk_{1}d\omega\nonumber
\end{align}
where $\tilde{f}$ and $\tilde{g}$ are smooth functions of $k_{1}$ and $\omega
$, related to $f$ and $g$ by the Fourier transform. \ The functions $f $ and
$g$ therefore cannot have compact support (though they may be chosen so as to
have arbitrarily small tails outside of a finite region).

The upshot of this example is that the ordinary theory of the massless scalar
field, formulated in terms of states specified on mixed spacelike and timelike
hypersurfaces, is one in which a natural generalization of \emph{SI} to the
case of fields (rather than particle states and detector settings) is
violated. \ I.e., the natural analogue of $P(\lambda|a,b)=P(\lambda)$ does not
hold. \ For example, consider disjoint compact regions $A,$ $B$ and $\Lambda$
on the initial data surface. Let $\lambda=(f(\Lambda),g(\Lambda))$ represent
the state the field in $\Lambda$, and let $a=(f(A),g(A))$ and $b=(f(B),g(B))$
represent the detector settings $a$ and $b$. Then it is the case that, given a
generic probability distribution on the space of initial data $f$ and $g$, the
probability of $\lambda$ (the restriction of $f$ and $g$ to region $\Lambda$)
will \emph{not} be independent of $a$ and $b.$ For example, if the functions
$f$ and $g$ vanish in the regions $A$ and $B$, then it must be the case that
they vanish in $\Lambda$ (otherwise $\Lambda$ would be a region in which $f$
and $g$ have compact support, which we know not to be the case).

\begin{figure}
[ptb]
\begin{center}
\includegraphics[
natheight=3.653800in,
natwidth=2.286600in,
height=3.7005in,
width=2.3272in
]%
{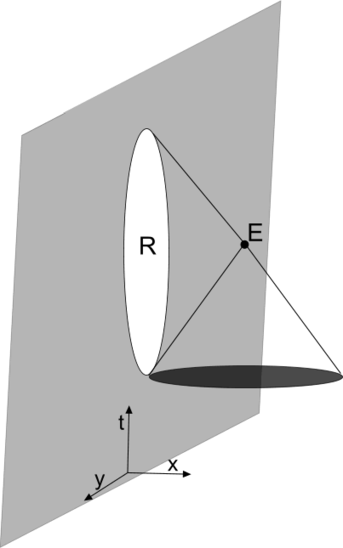}%
\caption{Compact timelike domain of dependence}%
\label{Fig2}%
\end{center}
\end{figure}

Note that, despite the failure of \emph{SI} in this context, we have perfectly
a well-posed initial value problem, and we even have compact domains of
dependence (see Figure \ref{Fig2}).%
%TCIMACRO{\FRAME{ftbpFU}{2.3272in}{3.7005in}{0pt}{\Qcb{Compact timelike domain
%of dependence}}{\Qlb{Fig2}}{timelike.png}%
%width 2.3272in;  height 3.7005in;  depth 0pt;  original-width 2.2866in;
%original-height 3.6538in;  cropleft "0";  croptop "1";  cropright "1";
%cropbottom "0";  filename 'Timelike.png';file-properties "XNPEU";}}}%
%BeginExpansion
%EndExpansion
Here, data in $R$ determines data in the region out to $E$. The nonlocal
constraint simply means that data in $R$ may not be specified freely.

\subsection{Timelike compactification}

\begin{figure}
[ptb]
\begin{center}
\includegraphics[
natheight=1.413100in,
natwidth=3.313100in,
height=1.4486in,
width=3.3598in
]%
{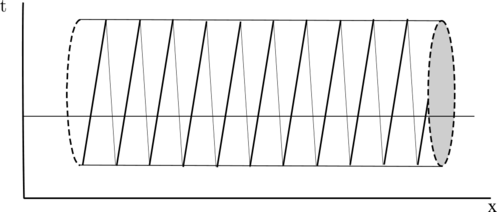}%
\caption{Timelike compactification}%
\label{CTC}%
\end{center}
\end{figure}

Consider once more the wave equation, this time in one space dimension (for
simplicity). \ And consider again the initial value problem, but on an
ordinary spacelike hyperplane. \ However, suppose that the spacetime on which
the field takes values is compactified in the time direction, so that the
entirety forms a cylinder (see Figure \ref{CTC}). This, too, is an example of
a theory whose initial value formulation possesses a nonlocal constraint.%
%TCIMACRO{\FRAME{ftbpFU}{3.3598in}{1.4486in}{0pt}{\Qcb{Timelike
%compactification}}{\Qlb{CTC}}{ctc.png}{\special{ language "Scientific Word";
%type "GRAPHIC";  maintain-aspect-ratio TRUE;  display "USEDEF";
%valid_file "F";  width 3.3598in;  height 1.4486in;  depth 0pt;
%original-width 3.3131in;  original-height 1.4131in;  cropleft "0";
%filename 'CTC.png';file-properties "XNPEU";}}}%
%BeginExpansion
%EndExpansion

The reason for the constraint is of course that the solutions must be
periodic. Whereas in the ordinary initial value problem, initial data may be
any smooth functions $f(x)=\phi(x,0)$ and $g(x)=\dot{\phi}(x,0)$, we now
require that $\phi(x,0)=\phi(x,T)$ and $\dot{\phi}(x,0)=\dot{\phi}(x,T)$,
where $T$ is the circumference of the cylinder. Solutions to the wave equation
can be written as sums of plane waves, with Fourier space representation%

\begin{equation}
\hat{\phi}(k,t)=\hat{F}(k)e^{-ikt}+\hat{G}(k)e^{ikt}\text{ ,}%
\end{equation}
Since these plane waves must have period $T$ (in the preferred frame dictated
by the cylinder), we have a constraint $k=\frac{2\pi n}{T}$ (where $n$ is a
positive or negative integer), so that initial data are no longer arbitrary
smooth functions of $k$
\begin{align*}
\hat{\phi}(k,0)  &  =\hat{F}(k)+\hat{G}(k)\\
\hat{\phi}_{t}(k,0)  &  =-ik(\hat{F}(k)-\hat{G}(k))
\end{align*}
but are rather constrained by the requirement $k=\frac{2\pi n}{T}$. Thus the
initial data are the functions%
\begin{align*}
\phi(x,0)  &  =\frac{1}{\sqrt{T}}%
%TCIMACRO{\dsum \limits_{n=-\infty}^{\infty}}%
%BeginExpansion
{\displaystyle\sum\limits_{n=-\infty}^{\infty}}
%EndExpansion
\hat{\phi}(\frac{2\pi n}{T},0)e^{i\frac{2\pi n}{T}x}dk\\
\phi_{t}(x,0)  &  =\frac{1}{\sqrt{T}}%
%TCIMACRO{\dsum \limits_{n=-\infty}^{\infty}}%
%BeginExpansion
{\displaystyle\sum\limits_{n=-\infty}^{\infty}}
%EndExpansion
\hat{\phi}_{t}(\frac{2\pi n}{T},0)e^{i\frac{2\pi n}{T}x}dk
\end{align*}
i.e., they consist of arbitrary sums of plane waves with wave number
$k=\frac{2\pi n}{T}.$

The restriction to a discrete (though infinite) set of plane waves means that
initial data do not have compact support; they are periodic in both space and
time. Thus,\emph{\ }as in the case of the mixed initial value problem, the
data cannot be specified freely. However, for sufficiently large $T$ or
sufficiently small $\Delta x$, the local physics is indistinguishable from the
local physics in ordinary Minkowski spacetime. Only at distance scales on the
order of $T$ does the compact nature of the direction become evident in the
repetition of the spatial structure.

\section{Contextuality}

The Kochen-Specker theorem points toward a kind of contextuality in quantum
mechanics, and indeed in any theory in which the properties of a system are
understood to be independent of the properties of other systems. The theorem
shows that, for systems described by quantum mechanics, the properties of
these systems cannot consistently be assigned values if the values respect a
certain seemingly natural criterion called `functional composition'
\cite{Red89}. Functional composition is the assumption that the value
$v(\hat{A}\hat{B})$ of an observable $\hat{A}\hat{B}$ which is the product of
commuting observables $\hat{A}$ and $\hat{B}$ is equivalent to the product
$v(\hat{A})v(\hat{B})$ of the values of each observable, as long as the
observables commute. Given this assumption, one can show that the following
set of operators, representing spin observables for a system composed of two
spin-$1/2$ particles, cannot simultaneously be assigned values in a way that
is consistent with the requirement that the values belong to the eigenvalue
spectrum of the operators%
\begin{equation}%
\begin{array}
[c]{ccc}%
I\otimes\hat{\sigma}_{z} & \hat{\sigma}_{z}\otimes I & \hat{\sigma}_{z}%
\otimes\hat{\sigma}_{z}\\
\hat{\sigma}_{x}\otimes I & I\otimes\hat{\sigma}_{x} & \hat{\sigma}_{x}%
\otimes\hat{\sigma}_{x}\\
\hat{\sigma}_{x}\otimes\hat{\sigma}_{z} & \hat{\sigma}_{z}\otimes\hat{\sigma
}_{x} & \hat{\sigma}_{y}\otimes\hat{\sigma}_{y}\text{ .}%
\end{array}
\end{equation}
Rather, the value assigned to a given observable must depend on whether it is
being measured along with the other (commuting) observables in its \emph{row},
or the other observables in its \emph{column.}

Recent work on generalizing the Kochen-Specker result to any theory admitting
an operational characterization shows that there is a sense in which any
theory that reproduces the predictions of quantum mechanics must be contextual
\cite{Spek05}. Without going into unnecessary detail, the general idea is that
any theory reproducing the predictions of quantum theory must be such that the
probabilities for various outcomes must in general depend on which other
properties are (simultaneously) measured. Such a result, however, is utterly
unsurprising in a theory with nonlocal constraints, so long as one recognizes
that the detector orientations \emph{themselves} are part of the system, since
the nonlocal constraint means that the degrees of freedom of the detectors are
not independent of those describing the particles. Indeed, from a
non-operational, closed-system point of view, one may view the contextuality
implicit in the Kochen-Specker theorem as \emph{implying} the existence of a
nonlocal constraint. This sheds light on the relationship between Bell's
theorem and the Kochen-Specker theorem, in that K-S essentially shows that any
local hidden-variable theory must violate statistical independence, while Bell
shows that any statistically independent theory must violate locality.

\section{Future directions}

Neither of the two examples above, examples of theories with nonlocal
constraints, appear to have any direct connection to quantum mechanics, though
the mixed initial value problem \emph{might} be so related. It is certainly
worth investigating what sort of theory emerges if one takes, for example, the
wave equation on three space and \emph{two} time dimensions (called an
`ultrahyperbolic' equation) and considers data on an initial Cauchy surface of
$3+1$ dimensions. \ Such a theory will also have nonlocal constraints, and
might give rise to interesting behavior when the extra time dimension is
averaged over in such a way as to generate an effective theory in $3+1$
spacetime dimensions.\ The obvious difficulty is that, if the extra time
dimension is not compact, there may be no obvious choice of measure over which
to average.

Perhaps more intriguingly, one might ponder the way in which the ordinarily
superfluous gauge degrees of freedom of modern gauge theories might serve as
nonlocal hidden variables. The vector potential in electrodynamics, for
example, ordinarily plays no direct physical role:\ only derivatives of the
vector potential, which give rise to the electric and magnetic fields,
correspond to physical \textquotedblleft degrees of freedom\textquotedblright%
\ in classical and quantum electrodynamics.\ The Aharonov-Bohm effect shows
that the vector potential does play an essential role in the quantum theory,
but the effect is still gauge-invariant.\footnote{One might also ponder the
connection with the closely related way in which energy enters into classical,
nongravitational physics. In the absence of gravity, only \emph{differences}
in energy are held to be observable, but when gravity enters the picture,
absolute values of energy are understood to be relevant. Or course, this in
turn leads to the cosmological constant problem when one attempts to couple a
quantum theory of matter to classical gravitation.} One might nevertheless
conjecture that there is an underlying theory in which the potential
\emph{does} play a physical role, one in which the physics is \emph{not}
invariant under gauge transformations.\footnote{'t Hooft has also gestured in
this direction - see \cite{tHo07}, p. 7.} The indeterminacy we associate with
quantum theory may then arise via epistemic limitations. More specifically, it
may be impossible for us to directly observe the vector potential, and the
uncertainties associated with quantum theory may arise from our ignorance as
to its actual (and nonlocally constrained) value. From this
perspective,quantum theory would be an effective theory which arises from
modding out over the gauge transformations.

Finally, recent work on decoherence and the emergence of classicality
\cite{SW08} suggests that the emergence of classicality requires very special
quantum states. For worlds with a large number of subsystems, hence a high
Hilbert space dimension, only a measure zero subset of the total set of
quantum states gives rise to distinctively classical behavior. \ Thus it seems
quite reasonable to suppose a similarly strong constraint on the states of a
hidden-variable theory.

\section{Conclusion}

The ideas sketched in the previous section are preliminary, of course, and
they are only two of many possible ways to construct theories which feature
nonlocal constraints. What the reader should take away from this paper, if
nothing else, is the idea that\ it is not all that difficult to construct
nonlocal theories which nevertheless local in the sense of being
Lorentz-invariant and not allowing superluminal signaling, and that such
theories are quite promising as deterministic or stochastic models of many of
the curious phenomena described by quantum mechanics.

\end{document}